\documentclass[proof]{WileyASNA-v1}% This produces two warnings related to the .cls file provided in the template: "You have requested document class `WileyASNA-v1', but the document class provides `WileyASNA-v2'." and "fixltx2e is not required after 2015(fixltx2e) All fixes are now in the LaTeX kernel."

\usepackage{amsmath}
\usepackage{braket}
\usepackage{mathtools}
\usepackage{tensor}
\usepackage{enumitem}
\usepackage{siunitx}

\DeclareSIUnit\lightspeed{\text{\ensuremath{c}}}
\DeclareSIUnit\boltzmannk{\text{\ensuremath{k_{\textup{B}}}}}

\newcommand*\idd[1]{\mathrm{d}#1\,}
\newcommand*\iddn[2]{\mathrm{d}^{#1}#2\,}

\newcommand*\CP{C\!P}

\DeclarePairedDelimiterX{\abs}[1]{\lvert}{\rvert}{#1}

\articletype{Original Article}%

\received{xx xx xxx}
\revised{xx xx xxxx}
\accepted{xx xx xxxx}

\begin{document}

\title{Non-inertial effects on \texorpdfstring{\(\CP\)}{CP}-violating systems}

\author[1]{V. M. G. Silveira*}

\author[1,2]{C. A. Z. Vasconcellos}

\author[1]{E. G. S. Luna}

\author[1]{D. Hadjimichef}

\authormark{V. M. G. SILVEIRA  et al. }

%\authormark{V. M. G. SILVEIRA \textsc{et al.}}

\address[1]{\orgdiv{Instituto de F\'{\i}sica}, \orgname{Universidade Federal do Rio Grande do Sul (UFRGS)}, \orgaddress{\city{Porto Alegre}, \state{Rio Grande do Sul}, \country{Brazil}}}

\address[2]{\orgname{International Center for Relativistic Astrophysics Network (ICRANet)}, \orgaddress{\city{Pescara}, \country{Italy}}}

\corres{*Av. Bento Gon\c{c}alves, 9500, Porto Alegre - RS, 91501-970, Brazil \email{viniciusmgsilveira@gmail.com}}

\abstract{An analysis of accelerated kaon decays based on the Unruh effect shows a slight decrease in a \(\CP\)~violation parameter for very high accelerations, as a consequence of the previously know increase in decay rates for non-inertial systems. Its consequences on the understanding of the relation between thermal and non-inertial phenomena are briefly discussed.}

\keywords{\(\CP\)~violation, Unruh effect, Matter--antimatter asymmetry, Kaons}

%\jnlcitation{\cname{\author{Silveira, V. M. G.} et al.} (\cyear{xxxx}), \ctitle{Non-inertial effects on \(\CP\)-violating systems}, \cjournal{Astron. Nachr.}, \cvol{xxxx;xx:x--x}.}

\maketitle

\section{Introduction}

 The phenomenon of  {\it CP violation} (\(\CP\)v) plays a crucial role in understanding the prevalence of matter over antimatter in our Universe. 
In Ref. 
\citep{Sakharov} it is shown that \(\CP\)v is  one of the necessary ingredients for baryogenesis, the principal mechanism for the existence of matter-antimatter asymmetry [see~\citep{Rubakov} for a review].

The\, existence of the  {\it Unruh effect} was established in  \citep{Fulling,Davies,Unruh},
which states that accelerated observers perceive the usual vacuum state of a quantum field theory (constructed around inertial observers) as a thermal bath of temperature
 \begin{equation}\label{TUnruh}
  T = \frac{\hslash a}{2\pi k_{\textup{B}}},
 \end{equation}
 where \(a\)~is the acceleration. A consequence of the Unruh effect is that proper decay rates of particles increase when they are accelerated, as one would expect for a particle in a thermal bath  \citep{Muller,VanzellaShort,VanzellaLong}.

 In this paper, we seek to show that \(\CP\)v is affected by non-inertial phenomena via an investigation of \(\CP\)v in the quark sector, which manifests itself in the Cabibbo--Kobayashi--Maskawa (CKM) matrix [see~\citep{Bigi} for a comprehensive presentation] and, e.g., in neutral kaon decays, the source of the first experimental detection of \(\CP\)v, reported in~\citep{Cronin}. \(\CP\)-violating observables can be obtained from the decay rates of the aforementioned kaons, which, in turn, vary when these particles are accelerated. This is summarized in the following heuristic equation,
 \begin{equation}
  \CP\text{v}= (\text{imbalance}) \times (\text{non-inertial modification}),
 \end{equation}
 where we seek to understand the non-inertial factor.

 Throughout this work we use natural units \(8\pi G = c = \hslash = k_{\textup{B}} = 1\), unless stated otherwise, and the \((-,+,+,+)\)~convention for the metric signature.

 A more detailed discussion of the research described in this text and related topics may be found in~\citep{Silveira}.

\section{\texorpdfstring{\(\CP\)}{CP} violation in the \texorpdfstring{\(K\)}{K}-system}\label{sec2}

 The existence of oscillations between \(K^0\) and \(\bar{K}^0\), coupled with the fact that both decay (weakly) in two pions,
 \begin{equation}
  K^{0}\longleftrightarrow 2\pi\longleftrightarrow \bar{K}^{0},
  \label{cp2}
 \end{equation}
 points to the conclusion that these are not independent states with respect to the weak interaction. Also of note is that these are not \(\CP\)~eigenstates,
 \begin{equation}
   \CP\ket {K^{0}} = -\ket {\bar{K}^{0}},\quad \CP\ket {K^{0}} = -\ket {\bar{K}^{0}}
  \label{cp3}
 \end{equation}
 while the two pions state is:
 \begin{equation}
  \CP\ket{\pi\pi} = \ket{\pi\pi}
 \end{equation}
 If the weak interaction is \(\CP\)-symmetric, its eigenstates must be the following,
 \begin{equation}
  \ket{K_{\substack{1\\2}}} = \frac{1}{\sqrt{2}}\bigl(\ket{K^{0}}\mp\ket{\bar{K}^{0}}\bigr),
 \end{equation}
 which, using Equation~\eqref{cp3}, can be seen to be \(\CP\)~eigenstates:
 \begin{equation}
  \CP \ket{K_{\substack{1\\2}}} = \pm\ket{K_{\substack{1\\2}}}.
  \label{cp6}
 \end{equation} 
 This is also clear if the following \(\CP\)~projector is defined,
 \begin{equation}
  \mathcal{P}_{\pm}=\frac{1}{2}(1\pm \CP),
  \label{cp6b}
 \end{equation}
 i.e., \(\mathcal{P}_{\pm} \ket{K_{\substack{1\\2}}} = \ket{K_{\substack{1\\2}}}\) and \(\mathcal{P}_{\pm}\ket{K_{\substack{2\\1}}} = 0\). It is then easy to see that, if the \(\CP\)~symmetry is not violated by the weak interaction, \(K_{1}\longrightarrow 2 \pi\) is the only permitted decay channel.

 Given the experimental detection of \(\CP\)v in weak processes, it follows that the weak eigenstates must be of the form
 \begin{align}
  \ket{K_{\textup{S}}}=\frac{1}{\sqrt{1+\abs{q}^2}}\bigl(\ket{K^{0}} - q\ket{\bar{K}^{0}}\bigr),\\
  \ket{K_{\textup{L}}}=\frac{1}{\sqrt{1+\abs{q}^2}}\bigl(\ket{K^{0}} + q\ket{\bar{K}^{0}}\bigr),
 \end{align}
 for some (complex) mixing parameter~\(q \neq 1\), since the \(q = 1\) case corresponds to~\(K_{1}\) and \(K_{2}\). These states are known as the  {\it small} (or  {\it short-lived}) kaon (\(K_{\textup{S}}\)) and 
the  {\it large} (or  {\it long-lived}) kaon (\(K_{\textup{L}}\)), since they differ with respect to their lifetimes (\(\tau_{\textup{S}} < \tau_{\textup{L}}\)) and masses (\(m_{K_{\textup{S}}} < m_{K_{\textup{L}}}\)).
 The measure of \(\CP\)v in \(K \longrightarrow 2 \pi\) decays is given by
 \begin{equation}
  \eta \coloneqq \frac{\mathcal{A}(K_{\textup{L}} \to\pi\pi)}{\mathcal{A}(K_{\textup{S}} \to\pi\pi)}
  \label{cp7d}
 \end{equation}
 i.e., the ratio between the amplitudes~\(\mathcal{A}\) of the channels.\footnote{This ratio is different depending on whether the decay products are neutral or charged pions (see the theoretical justification in~\citep{Bigi} and the data available in~\citep{PDG}), but for either case the results are qualitatively the same.\label{diffpi}} The relation between~\(\eta\) and~\(q\) can be obtained from
 \begin{equation}
  \eta = \frac{\braket{\pi\pi\vert\mathcal{P}_{+}\vert K_{\textup{L}}}}{\braket{\pi\pi\vert\mathcal{P}_{+}\vert K_{\textup{S}}}} = \frac{1-q}{1+q}.
  \label{cp7c}
 \end{equation}
 The value of the amplitude of \(\eta\) is approximately~\(\num{2e-3}\) [see~\citep{PDG} and Footnote~\ref{diffpi}]. 
A useful expression for \(\abs{\eta}^{2}\) in terms of the decay rates~\(\Gamma\), to be used below, is
 \begin{equation}
  \abs{\eta}^2 = \frac{\Gamma(K_{\textup{L}}\to\pi\pi)}{\Gamma(K_{\textup{S}}\to\pi\pi)}.
  \label{eta2}
 \end{equation}

\section{Non-inertial effects and kaon
  decays}\label{decayaccel}

 The results of~\citep{Muller} show that scalar fields decay at a higher rate when accelerated---specifically, that their decay rate with respect to the proper time increases with the acceleration---while \citep{VanzellaShort,VanzellaLong} extend this conclusion to spinorial fields. This phenomenon is best understood in light of the Unruh effect [see~\citep{Crispino} for a review], which relates the vacuum state of a quantum field theory, as described by an inertial observer, to a thermal (KMS) state, as described by an accelerated observer, with temperature given in Equation~\eqref{TUnruh}. An inertial observer would then conclude that an accelerated particle decays faster due to the increase in its kinetic energy, while a co-accelerated observer would perceive an interaction between the particle and the thermal bath.

 By using a similar model to the one introduced in~\citep{Muller}, we compute the dependency of the decay rates \(\Gamma(K \longrightarrow \pi\pi)\) on the acceleration and, with the use of Equation~\eqref{eta2}, analyse the behaviour of~\(\eta\) when the kaons are accelerated.

 \subsection{A model for accelerated decays}
  We model a kaon species as a scalar\footnote{While kaons are pseudoscalar fields, we assume they can be well modeled by the simpler scalar fields.} field~\(K\) with mass~\(m_{K}\) and the decay products, two pions, as scalar fields~\(\pi_{1}\) and~\(\pi_{2}\) with mass~\(m_{\pi}\). Their interaction is described by the following Lagrangian,
  \begin{equation}
   \mathcal{L}_{\textup{I}}(x) = G_{\Gamma}K(x)\pi_{1}(x)\pi_{2}(x),
   \label{un1}
  \end{equation}
  where \(G_{\Gamma}\)~is a coupling parameter. The transition amplitude can be computed perturbatively and is given, up to first order, by
  \begin{equation}
   \begin{split}
    \mathcal{A}(\mathbf{k}_{1},\mathbf{k}_{2}) &= \bra{\mathbf{k}_{1},\mathbf{k}_{2}}\otimes\bra{0}S\ket{i}\otimes\ket{0}\\
    &= G_{\Gamma}\int\iddn{4}{x}
    \braket{0 \vert K(x) \vert i}\prod_{j=1}^{2}\braket{\mathbf{k}_{j} \vert \pi_{j}(x) \vert 0}.
   \end{split}
   \label{un2}
  \end{equation}
  We assume that the initial state~\(\ket{i}\) of the kaon corresponds to an oscillation mode~\(f(x)\) and that the final states of the pions have definite momenta~\(\mathbf{k_{1}}\) and~\(\mathbf{k_{2}}\). Taking the square of the absolute value of the amplitude and integrating over the momenta, we obtain the decay probability for this process:
  \begin{equation}
   \mathcal{P} = G_{\Gamma}^{2}\int\iddn{4}{x}\iddn{4}{x^{\prime}} f^{*}(x)f(x^{\prime})\prod_{j=1}^{2}\braket{0 \vert \pi_{j}^{\dagger}(x)\pi_{j}(x^{\prime}) \vert 0},
   \label{un3}
  \end{equation}
  Evaluating this expression requires the following form of the Wightman function of a scalar field~\(\pi\) in terms of the spacetime interval~\(\Delta s\) of the two events~\(x\) and~\(x^{\prime}\),
  \begin{equation}
   \braket{0 \vert \pi^{\dagger}(x)\pi(x^{\prime}) \vert 0} = i\frac{m}{8\pi}\frac{H_{1}^{(2)}(m\Delta s)}{\Delta s},
   \label{un4}
  \end{equation}
  where \(H_{1}^{(2)}\)~is a Hankel function of the second kind.

  To account for the kaon's acceleration, we compute~\(\Delta s\) over a uniformly accelerated trajectory,
  \begin{equation}
   \Delta s^{2} = -\frac{4}{a^{2}}\sinh^{2}\Bigl(\frac{a}{2}\bigl(\tau-\tau^{\prime}\bigr)\Bigr),
   \label{un9}
  \end{equation}
  where~\(\tau\) and~\(\tau^{\prime}\) are the proper times of the events~\(x\) and~\(x^{\prime}\), respectively, and \(a\)~is the proper acceleration. Since we assume that the kaon follows a well defined trajectory, we can take~\(f\) to be peaked over it, i.e.,
  \begin{equation}
   f(x) = h(\mathbf{x}(\tau))e^{-iM\tau}
   \label{un5}
  \end{equation}
  in the instantaneous rest frame. The expression for the decay probability then becomes
  \begin{equation}
   \mathcal{P} = -\frac{G_{\Gamma}^{2}\kappa}{64\pi^{2}}m^{2}\int_{-\infty}^{\infty}\int_{-\infty}^{\infty}\idd{\tau}\idd{\tau^{\prime}}e^{iM(\tau-\tau^{\prime})}\frac{\bigl[H_{1}^{(2)}(m\Delta s)\bigr]^{2}}{\abs{\Delta s^{2}}},
   \label{un6}
  \end{equation}
  where \(\kappa\)~is given by \(\kappa =\abs{\int\iddn{3}{x}h(\mathbf{x})}^2\). A change of variables allows the decay rate (i.e., the decay probability by unit of proper time) to be written as
  \begin{equation}
   \Gamma = -\frac{G_{\Gamma}^{2}\kappa}{64\pi^{2}}m^{2}\int_{-\infty}^{\infty}\idd{v} e^{iMv}\frac{\bigl[H_{1}^{(2)}(m\Delta s)\bigr]^{2}}{\abs{\Delta s^{2}}},
   \label{gammav}
  \end{equation}
  or, using yet another change of variables and Equation~\eqref{un9},
  \begin{equation}
   \Gamma = -\frac{G_{\Gamma}^{2}\kappa}{128\pi^{2}}m^{2}a\!\int_{-\infty}^{\infty}\idd{u}e^{i2Mu/a}\frac{\bigl\{H_{1}^{(2)}[2m\sinh(u)/a]\bigr\}^{2}}{\sinh^{2}(u)}.
   \label{gamma}
  \end{equation}

 \subsection{Results and Analysis}
\begin{figure}[t]
   \includegraphics[width=\linewidth]{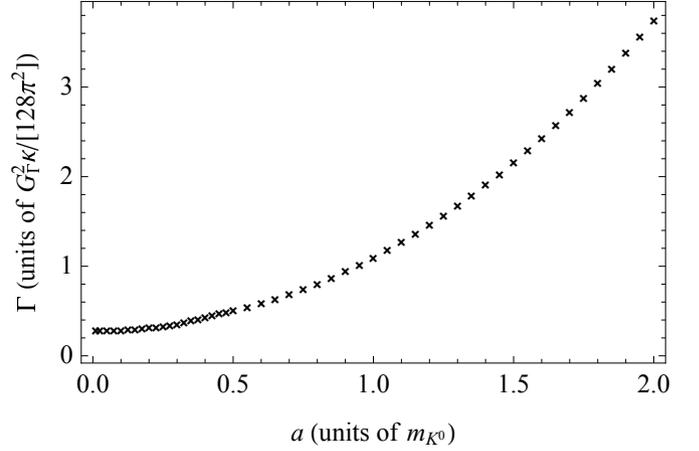}
   \caption{\label{plotgamma}Numerical results for the decay rate~\(\Gamma\) as a function of the acceleration~\(a\).}
  \end{figure}

  Determining the behaviour of~\(\Gamma\) with respect to~\(a\) requires the computation of the singular and highly oscillatory integral in Equation~\eqref{gamma}. To this end, we split this integral into its singular part (treated analytically) and non-singular part (treated numerically). The values for~\(m_{K}\) and~\(m_{\pi}\) can be found in Table~\ref{tabmass}, in units of~\si{\mega\eV\!\per\lightspeed\squared} and of~\(m_{K^{0}}\), with the correspondence \(m_{K} = m_{K^{0}}\) and \(m_{\pi} = m_{\pi^{0}}\) (see Footnote~\ref{diffpi}). The results of the computation can be seen in Figure~\ref{plotgamma}, which plots~\(\Gamma\) (in units of~\(G_{\Gamma}^{2}\kappa/(128\pi^{2})\)) against~\(a\) (in units of \(m_{K^{0}}\)).
  \begin{table}[b]
   \centering
   \caption{\label{tabmass}Values for the masses of the pion~\(\pi^{0}\) and the neutral kaon~\(K^{0}\), and the mass difference between \(K_{\textup{L}}\)~and~\(K_{\textup{S}}\), given in~\citep{PDG}.}
   \begin{tabular*}{\columnwidth}{c@{\extracolsep{\fill}}S[table-format = 3.5(2)e+2]S[table-format = 1.6e+2]}
    \toprule & \multicolumn{1}{c}{\si{\mega\eV\!\per\lightspeed\squared}} &  \multicolumn{1}{c}{\(m_{K^{0}}\)}\\
    \midrule
    \(m_{K^{0}}\) & 497.611(13) & 1\\
    \(m_{\pi^{0}}\) & 134.9770(5) & 0.271250\\
    \(m_{K_{\textup{L}}}-m_{K_{\textup{S}}}\) & 3.484(6)e-12 & 7.001e-15\\
    \bottomrule
   \end{tabular*}
  \end{table}

  It is clear that \(\Gamma\)~increases with an increase in~\(a\), in accordance with~\citep{Muller,VanzellaShort,VanzellaLong}, but the scales at which theses effects may be detected are unreachable with current experimental capabilities. An acceleration on the same scale of the kaon mass (in natural units) is of the order of~\SI{e32}{\meter\per\second\squared}, many orders of magnitude higher than the highest accelerations obtained by laser accelerators such as the Texas Petawatt laser, reaching no more than~\SI{e23}{\meter\per\second\squared} 
[see the data reported in~\citep{TPLaser}]. That \(\CP\)v manifests itself in {neutral} kaon systems also presents a problem when it comes to accelerating particles, though (given the fundamental conclusions that can be drawn from the Unruh effect) thermal phenomena of a different nature may allow for the same conclusions.

  The dependence of the \(\CP\)-violating observable~\(\eta\) on the value of the acceleration can be analysed by combining the results obtained for~\(\Gamma\) with Equation~\eqref{eta2} and noting that the non-negligible difference in the masses of~\(K_{\textup{S}}\) and~\(K_{\textup{L}}\) (whose value is given in Table~\ref{tabmass}) influences the growth rate of~\(\Gamma\). Thus, we compute \(\Gamma(K_{\textup{S}}\to\pi\pi)\) using \(m_{K_{\textup{S}}} \approx m_{K^{0}}\), as above, and \(\Gamma(K_{\textup{L}}\to\pi\pi)\) using \(m_{K_{\textup{L}}} \approx m_{K^{0}} + (m_{K_{\textup{L}}} - m_{K_{\textup{S}}})\). Care must be taken, however, since the coupling constants \(G_{\Gamma}(K_{\textup{S}}\to\pi\pi)\) and \(G_{\Gamma}(K_{\textup{L}}\to\pi\pi)\) must differ considerably in order for~\(\abs{\eta} \approx \num{2e-3}\). To address this in our computation, we consider the quantity~\(\eta^{\prime}\), given by
  \begin{equation}
   \eta = \beta\eta^{\prime}, \qquad \beta = \frac{G_{\Gamma}(K_{\textup{L}}\to\pi\pi)}{G_{\Gamma}(K_{\textup{S}}\to\pi\pi)}.
  \end{equation}

\begin{figure}[t]
   \includegraphics[width=\linewidth]{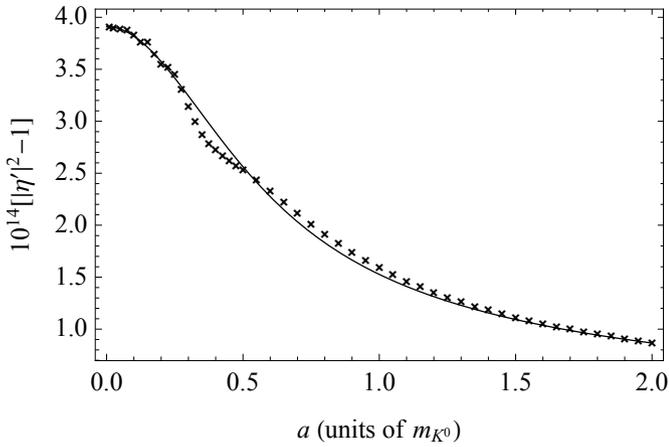}
   \caption{\label{ploteta}Numerical results for~\(\abs{\eta^{\prime}}^{2}-1\) as a function of the acceleration~\(a\) and plot of an approximation obtained from singular integrals.}
  \end{figure}

  Figure~\ref{ploteta} shows the plot of~\(\abs{\eta^{\prime}}^{2}-1\) against~\(a\), alongside the graph of function approximating it, obtained by computing the singular parts of the integrals involved in the computation of the decay rates. These plots indicate that~\(\abs{\eta}^{2}\) decreases with higher values of~\(a\), though the amplitude of its variation is of the order of~\(\num{e-13}\abs{\eta}^{2}\), out of reach of experimental investigations (the current experimental uncertainty for~\(\abs{\eta}\) is of at least~0.4\%). This still implies that \(\CP\)v is influenced by non-inertial effects, in fact, \(\CP\)-violating processes seem to be slightly less efficient at very high accelerations.

\section{Conclusions}

 The investigation on \(\CP\)v in accelerated systems delineated above allows for the conclusion that non-inertial effects influence \(\CP\)-violating processes by decreasing the degree to which they break this symmetry. The computation of the dependency of the decay rates on the acceleration are in accordance with those obtained in~\citep{Muller} and allow for analysis of the behaviour of the \(\CP\)-violating observable~\(\eta\), which has been show to decrease by around three parts in \num{e14} for an acceleration of the order of \(2m_{K_{0}} \approx \SI{4e32}{\meter\per\second\squared}\).

 In order to further the understanding of phenomena of the nature described above, some questions may be posed:
 \begin{enumerate}
  \item Is this behaviour also present in other \(\CP\)-violating systems?\label{CPq}
  \item To what degree are non-inertial and thermal effects analogous (given the Unruh effect)?\label{Thermq}
 \end{enumerate}
 The framework of this paper may be readily extended to other (pseudo)scalar fields that present a difference in the masses of their weak eigenstates, providing a partial answer to Question~\ref{CPq}. \citep{Cozzella}~investigates non-inertial effects on neutrino mixing, also related to \(\CP\)v and, thus, to Question~\ref{CPq}. With respect to Question~\ref{Thermq}, \citep{Brauner}~presents results on thermal effects on \(\CP\)-violating observables in the quark sector, which can serve as a starting point towards an answer. This answer, in turn, may shed light on the compatibility between computations involving thermal fields in inertial reference frames and those describing accelerated fields.

\section*{Acknowledgments}

 The authors would like to thank George E. A. Matsas for enlightening comments. The research was supported by Conselho Nacional de Desenvolvimento Cient\'{\i}fico e Tecnol\'ogico (CNPq) and Universidade Federal do Rio Grande do Sul (UFRGS).

\bibliography{Bib.bib}

\end{document}